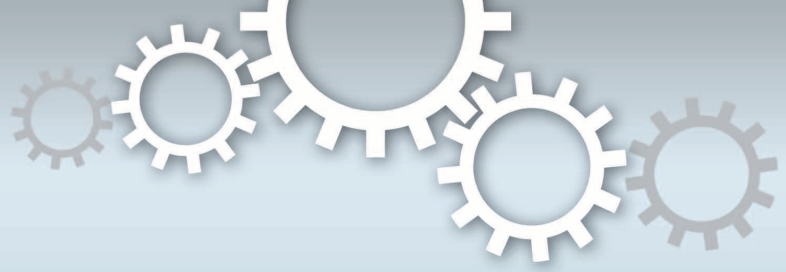

OPEN

# The phase diagram and hardness of carbon nitrides

Huafeng Dong[1,2], Artem R. Oganov[1,2,3,4], Qiang Zhu[1,2] & Guang-Rui Qian[1,2]

[1]Department of Geosciences, Stony Brook University, Stony Brook, New York 11794-2100, U.S.A., [2]Center for Materials by Design, Institute for Advanced Computational Science, Stony Brook University, Stony Brook, New York 11794-2100, U.S.A., [3]Moscow Institute of Physics and Technology, 9 Institutskiy Lane, Dolgoprudny City, Moscow Region 141700, Russia, [4]School of Materials Science, Northwestern Polytechnical University, Xi'an 710072, China.

Novel superhard materials, especially those with superior thermal and chemical stability, are needed to replace diamond. Carbon nitrides (C-N), which are likely to possess these characteristics and have even been expected to be harder than diamond, are excellent candidates. Here we report three new superhard and thermodynamically stable carbon nitride phases. Based on a systematic evolutionary structure searches, we report a complete phase diagram of the C-N system at 0–300 GPa and analyze the hardest metastable structures. Surprisingly, we find that at zero pressure, the earlier proposed graphitic-$C_3N_4$ structure ($P\bar{6}m2$) is dynamically unstable, and we find the lowest-energy structure based on s-triazine unit and s-heptazine unit.

The carbon-nitrogen (C-N) system was long believed to have materials harder than diamond[1]. Recently, carbon nitrides attracted attention due to their potential applications in photocatalysis[2], photodegradation[3] and photoelectrochemical anticorrosion[4] technology. However, studies of carbon nitrides under pressure face a big problem: neither from theory, nor from experiment it is clear which compositions (i.e., which C/N ratios) will be stable at high pressure, and which compositions will have optimal properties, such as hardness.

Experiments face challenges related to metastability, selection of precursors, determination of the crystal structures and chemical compositions from tiny samples[5–7]; theoretical calculations suffer from assumptions of certain stoichiometries, e.g., $C_3N_4$[1], $CN$[8], $C_2N$[9,10], $CN_2$[11], $C_{11}N_4$[12], $C_3N_2$[13], $CN_6$.... Some of them are meaningful, while most of them are probably not.

It is difficult to solve this problem because numerous compositions must be tested at each pressure point. For example, at $P = 0$ GPa, one needs to test compositions of 1 : 1, 1 : 2, 1 : 3,...; 2 : 1, 3 : 1, 4 : 1,...; 2 : 3, 2 : 5, 2 : 7,...... When the pressure changes, e.g. $P = 5, 10, 20...$ GPa, this needs to be re-checked. Therefore, it becomes a major effort.

To solve this problem, we used the *ab initio* evolutionary algorithm USPEX[14–17], which can simultaneously find stable stoichiometries and the corresponding structures in multicomponent systems. First, we carried out variable-composition calculations at pressures of 1 atm, 30, 60, 80, 100, 150, 200, and 300 GPa to find the stable and nearly stable compositions (found compositions: $C_{11}N_4$, $C_2N$, $CN$, $C_3N_4$, $CN_2$, $CN_6$). Then, for each of these compositions, we performed fixed-composition calculations with different numbers of formula units at different pressures. We not only obtained the stable compositions and structures at each pressure, we also analyzed the hardness of all the stable and metastable structures and found out the hardest structures and compositions, which provide a solid basis for future synthesis of promising ultrahard materials, stable or metastable. Moreover, we have three other novel findings:

1. We uncovered three new superhard phases which are more stable than previous proposals.
2. Graphitic-$C_3N_4$ (space group: $P\bar{6}m2$, based on s-triazine unit)[18–20] was reported to be stable at ambient pressure and has rich potential applications in photocatalysis[2], photodegradation[3] and photoelectrochemical anticorrosion[4]. However, we found graphitic-$C_3N_4$ is not dynamically stable and we uncovered two new structures based on s-triazine unit and s-heptazine unit, which are more stable at ambient pressure.
3. We determined a complete pressure-composition phase diagram of the C-N system at 0–300 GPa, which provides basis to guide the future experimental synthesis of superhard C-N materials.

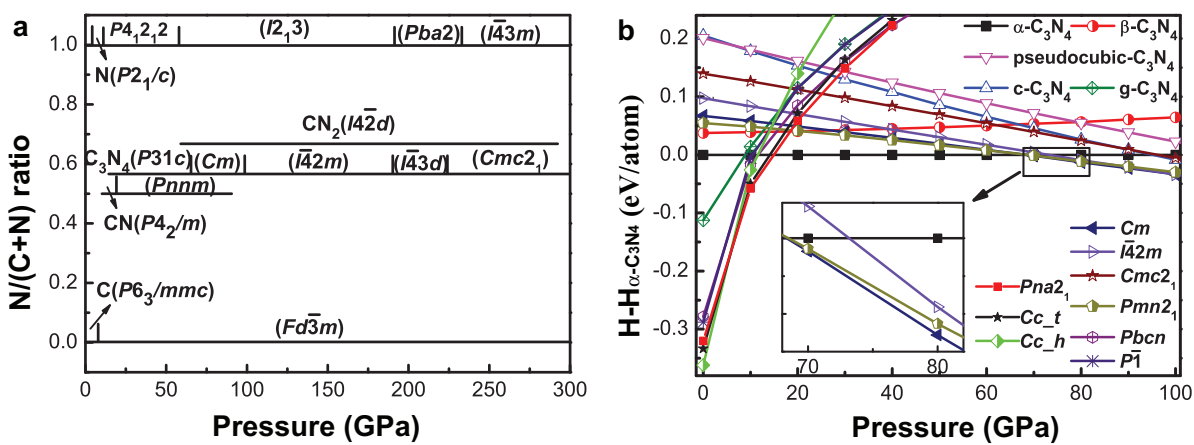

**Figure 1** | (A) Pressure-composition phase diagram of the C-N system. (B) Enthalpy curves (relative to α-$C_3N_4$) of the five earlier proposed structures[18] and the newly predicted structures.

## Results

**Phase diagram.** Detailed enthalpy calculations for the most stable structures allowed us to reconstruct the pressure-composition phase diagram (Fig. 1). The first thermodynamically stable carbon nitride, $P4_2/m$-CN, appears at the pressure of just 14 GPa. This is a superhard 3D-polymeric structure, as all the other stable carbon nitrides. The predicted phase diagram indicates that $I\bar{4}2d$-$CN_2$ is stable at 59–298 GPa; $P31c$-$C_3N_4$ (i.e., α-$C_3N_4$) at 22–68 GPa; $Cm$-$C_3N_4$ at 68–98 GPa (reported for the first time); $I\bar{4}2m$-$C_3N_4$ at 98–187 GPa (reported for the first time); $I\bar{4}3d$-$C_3N_4$ (i.e., cubic-$C_3N_4$) at 187–231 GPa; $Cmc2_1$-$C_3N_4$ at 224–300 GPa (reported for the first time); $P4_2/m$-CN at 14–22 GPa; and $Pnnm$-CN, stable at 22–97 GPa (see Supplementary Table S1 online for details of structure detailed structural information). These stable structures have lower free energy than any isochemical mixture of other compounds or pure elements (see Supplementary Fig. S1 online). Their main features are the presence of only single C-C, C-N and N-N bonds, with fourfold (tetrahedral) coordination of all C atoms and threefold coordination of all N atoms. For all the stable structures, we computed phonons (see Supplementary Fig. S2 online) and elastic constants (see Supplementary Table S1 online) at ambient pressure, and found them to be dynamically and mechanically stable. Their band structures exhibit wide band gaps as a consequence of strongly localized electrons (see Supplementary Fig. S3 online). Remarkably, all of these phases have three-dimensional frameworks of short covalent bonds, which are responsible for their extreme hardness (Table 1).

$CN_2$ is stable in the tetragonal $I\bar{4}2d$ structure at 59–298 GPa (Fig. 2a)[11]. In this structure, each C atom is tetrahedrally bonded with four N atoms and each N atom is three-coordinate (2 C-N bonds and 1 N-N bond). Notably, under normal conditions, the $I\bar{4}2d$-$CN_2$ structure has an atomic number density of 0.167 atoms/Å$^3$, which is just a little lower than that of diamond (GGA result: 0.176 atoms/Å$^3$; experimental result: 0.178 atoms/Å$^3$ [18]) (Table 1). Moreover, its N-N bond length is only 1.359 Å, and C-N bond length is 1.480 Å. Both are shorter than the C-C bond length of diamond (1.547 Å) (Table 1).

$C_3N_4$ has attracted much attention for more than 20 years[1,18]. It is not only because the predicted bulk modulus of the cubic-$C_3N_4$ structure is higher than that of diamond[21,22], but also because of the rich potential applications of the graphitic-$C_3N_4$ in water splitting, organic photosynthesis and environmental remediation[23]. The two most important graphitic carbon nitrides discussed in the literature are based on s-triazine[24] and s-heptazine (i.e. tri-s-triazine)[25] units. The graphitic-$C_3N_4$ (space group: $P\bar{6}m2$, based on s-triazine unit) was reported to be stable at ambient conditions, and there are numerous reports in the literature that approach the synthesis of this material[18–20]. Besides, many theoretical studies are based on this structure[26]. However, we found it to be dynamically unstable (see Supplementary Fig. S2 online). This is in agreement with the prior prediction of Deifallah et al.[27] and Bojdys et al.[28], which reported a buckled-graphitic-$C_3N_4$ to be more stable than the planar graphitic-$C_3N_4$. In addition, we found the most stable structures based on s-triazine unit or s-heptazine unit are not just a buckled structure, all their layers are connected by one nitrogen atom with sp$^2$ C-N bonds[29], as shown in Supplementary Fig. S4. We found that the lowest-energy form of structures based on s-triazine unit and s-heptazine unit are almost the same. They have similar topology and belong to the same space group $Cc$ (No. 9). The s-heptazine-based structures were postulated on the basis of density-functional calculations[25] and experiments[28] to be more stable at ambient conditions. Our predictions support this conclusion: $Cc$-$C_3N_4$ (s-heptazine) is lower in energy than $Cc$-$C_3N_4$ (s-triazine). Moreover, we have compared their energy with other theoretical proposals[25,29], as shown in Supplementary Table S5 online. The newly found structures have lower energies than the previously proposed structures. We must emphasize that all calculations on zero-pressure phases of $C_3N_4$ took account of van der Waals interactions (see Methods).

At 22–68 GPa, the α-$C_3N_4$ structure is the most stable. The α-$C_3N_4$ structure is a three-dimensional framework, where each C atom has four inequivalent bonds to N atoms, and each N atom has three inequivalent bonds to C atoms (Fig. 2b). This kind of framework can very easily lead to an asymmetric charge distribution and weak piezoelectricity, which has been confirmed by our calculations (see Supplementary Table S2 online) and recent experimental work[30]. At 68–98 GPa, the $Cm$-$C_3N_4$ structure is more stable than any other known $C_3N_4$ polymorph, and also consists of corner-sharing $CN_4$ tetrahedron (Fig. 2c). We also found another new structure at the same pressure range, $Pmn2_1$-$C_3N_4$ (see Supplementary Fig. S4 online), which is just 2 meV/atom higher in enthalpy than the $Cm$-$C_3N_4$ structure. The $I\bar{4}2m$-$C_3N_4$ structure (Fig. 2d) is stable at 98–187 GPa. However, at pressures above 187 GPa this structure gives way to a high-density structure, cubic-$C_3N_4$, which is denser and less compressible (i.e. has higher bulk modulus) than diamond. Besides, it has the largest shear modulus and Young's modulus among all stable carbon nitrides (Table 2). At > 224 GPa (and at least to 300 GPa), the orthorhombic $Cmc2_1$-$C_3N_4$ structure, which has not been reported before, is stable. Its structure (Fig. 2f) can be described as an ABAB sequence of puckered graphitic-$C_3N_4$ layers with strong covalent bonds formed between the layers.

Table 1 | Hardness (GPa), computed by microscopic models, enthalpy of formation (EF) (eV/atom), and atomic density (atoms/Å³) for diamond and stable C-N phases at zero pressure

| Structures | EF | density | Hardness (GPa) | | | |
| --- | --- | --- | --- | --- | --- | --- |
| | | | Oganov | Šimůnek | Gao | Others |
| Diamond | | 0.176 | 89.2 | 90.7 | 93.0 | 93.6[a] |
| $I\bar{4}2d$—$CN_2$ | 0.813 | 0.167 | 85.6 | 89.0 | 82.2 | 77.4[b] |
| $P31c$-$C_3N_4$ | 0.504 | 0.161 | 78.1 | 81.3 | 72.0 | 82.7[c] |
| $Cm$-$C_3N_4$ | 0.571 | 0.165 | 75.5 | 82.0 | 71.7 | |
| $I\bar{4}2m$—$C_3N_4$ | 0.602 | 0.167 | 80.2 | 83.0 | 70.5 | |
| $I\bar{4}3d$—$C_3N_4$ | 0.710 | 0.173 | 83.8 | 86.8 | 75.1 | 92[c] |
| $Cmc2_1$-$C_3N_4$ | 0.644 | 0.167 | 79.1 | 76.1 | 72.7 | |
| $P4_2/m$-CN | 0.395 | 0.155 | 58.3 | 70.5 | 54.7 | |
| $Pnnm$-CN | 0.422 | 0.160 | 59.6 | 72.7 | 57.0 | 62.3[d] |

[a]Reference[37]. [b]Reference[11]. [c]Reference[21]. [d]Reference[31].

At 14–22 GPa, CN is stable in the $P4_2/m$-CN structure. When pressure is above 22 GPa, the $Pnnm$-CN structure is more stable (22–97 GPa), which is consistent with previous theoretical predictions[8,31]. The $P4_2/m$-CN structure has a three-dimensional network of covalent bonds composed of tetragonal C-N rings, connected by the C-C bonds (Fig. 2g), while the $Pnnm$-CN structure is composed of strongly puckered graphene layers of composition CN, connected to each other by C-C bonds (Fig. 2h). Both the $P4_2/m$-CN and $Pnnm$-CN structures have four-coordinate C atoms, connected with 3 N atoms and 1 C atom; and three-coordinated N, connected with 3 C atoms. The C-C bond lengths in the $P4_2/m$-CN and $Pnnm$-CN structures are 1.585 Å and 1.605 Å, respectively, i.e. longer than in diamond (1.547 Å). Interestingly, the $P4_2/m$-CN structure is similar to the host-guest structure recently predicted as a metastable form of carbon[32].

Komatsu[9] reported synthesis of sp³-bonded carbon nitride $C_2N$, but he did not provide the detail structural information. We have tried to find $C_2N$ structures based on Komatsu's report, but all the structures that we found are not thermodynamically stable under pressure. Horvath-Bordon, Kroke, McMillan et al.[33], showed a nice work on the synthesis of crystalline carbon nitride imide phase, C2N2(NH) under high pressure and high temperature conditions, which indicated that hydrogen does help carbon nitrides to be stable and studying the ternary C-N-H system should be interesting.

**Hardness.** Hardness is one of the biggest factors that stimulated interest in carbon nitrides. To study their hardness, we have used the Oganov[34,35], Šimůnek[36], and Gao[37] models that are based on microscopic parameters, and the Chen model[38] which is based on macroscopic parameters. All the results, including the Voigt-Reuss-Hill elastic moduli, are listed in Tables 1 and 2.

The hardest carbon nitride among the stable structures is $I\bar{4}2d$-$CN_2$, based on all microscopic models. According to the hardness models of Oganov, Šimůnek, and Gao, its hardness is 85.6, 89.0 and 82.2 GPa, respectively. To our surprise, its hardness is very close to that of diamond (89.2, 90.7 and 93.0 GPa according to Oganov, Šimůnek, and Gao model). Moreover, its hardness is higher than that of cubic-$C_3N_4$, which was believed to be the hardest carbon nitride. Based on the microscopic models all the stable carbon nitrides are superhard. In addition, the three micro-models give similar values of hardness for every predicted structure, as can be seen in Table 1, because in all these models the hardness is determined by bond lengths and bond strengths, even though the models do differ mathematically and ideologically.

Based on the macroscopic Chen model, the hardest structure is $Cmc2_1$-$C_3N_4$ (Table 2). The hardness value in this model is determined by parameter $k^2G$ ($k = G/B$, $G$ and $B$ are shear and bulk moduli, respectively). We found that the $Cmc2_1$-$C_3N_4$ structure has the largest $k^2G$ among the stable structures (Table 2). This makes the $Cmc2_1$-$C_3N_4$ structure the hardest one among the stable structures according to the Chen model. Notably, the predicted Poisson's ratio of the $Cmc2_1$-$C_3N_4$ structure is 0.1348, almost the lowest among all the stable structures ($\nu = 0$ means that the material will not deform in a direction perpendicular to the applied load). According to the Chen model, all the stable structures are superhard, which is consistent with the prediction of the microscopic models. While the exact hardness values produced by the microscopic and macroscopic models are in this case rather different (which is unusual), the qualitative conclusions are similar. Note, on passing, that the bulk modulus of cubic-$C_3N_4$ is larger than that of diamond, in good agreement with theoretical studies[21,22].

Metastable structures can often be synthesized by choosing appropriate precursors and/or controlling conditions such as the quench rate[20]. If such a metastable phase has superior properties and can be depressurized to ambient conditions, it will be attractive for applications. For example, the hardest structure does not have to be the most stable one. For each structure/composition, its distance from the thermodynamic convex hull is the appropriate measure of its (meta)stability: the closer it is to the convex hull, the more stable it is. To find the hardest structure and the hardness-favored compositions in the C-N system, we also calculated the hardness and mechanical properties of metastable structures in a reasonable range of enthalpies (we have analyzed all structures within 0–0.2 eV/atom from the convex hulls at all pressures in the range 0–300 GPa). The five hardest structures in the C-N system, selected using the Oganov and Chen models, and their hardnesses, are listed in Supplementary Table S3 online. All of them are dynamically and mechanically stable at ambient conditions based on the computed phonon dispersions and elastic constants (see Supplementary Fig. S5 and Table S4 online).

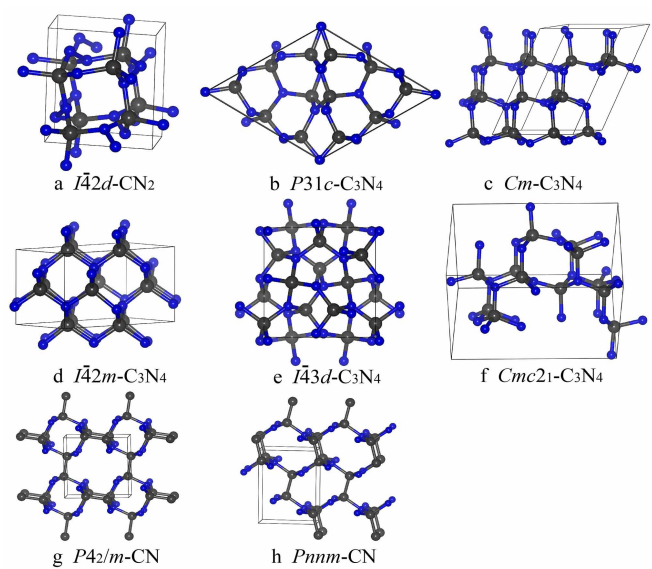

**Figure 2 | Crystal structures of carbon nitrides.** (A) $I\bar{4}2d$-$CN_2$. (B) $P31c$-$C_3N_4$. (C) $Cm$-$C_3N_4$. (D) $I\bar{4}2m$-$C_3N_4$. (E) $I\bar{4}3d$-$C_3N_4$. (F) $Cmc2_1$-$C_3N_4$. (G) $P4_2/m$-CN. (H) $Pnnm$-CN. Grey (large) and blue (small) spheres denote C and N atoms, respectively. $I\bar{4}2m$-$C_3N_4$ structure is a vacancy-ordered derivative of the diamond structure.

Table 2 | Hardness (GPa), computed by the macroscopic Chen model, bulk modulus B (GPa), shear modulus G (GPa), Young's modulus E (GPa), $k^2G$ ($k = G/B$) and Poisson's ratio $v$ for diamond and the stable C-N phases at zero pressure

| Structures | B | G | E | $k^2G$ | $v$ | $H_{Chen}$ |
|---|---|---|---|---|---|---|
| Diamond | 434.9 | 520.6 | 1116.4 | 748.4 | 0.0722 | 92.9 |
| $I\bar{4}2d$−CN$_2$ | 398.3 | 351.5 | 814.9 | 274.1 | 0.1589 | 50.3 |
| $P31c$-C$_3$N$_4$ | 383.6 | 330.6 | 770.4 | 245.5 | 0.1653 | 47.0 |
| $Cm$-C$_3$N$_4$ | 353.6 | 328.3 | 752.1 | 283.0 | 0.1455 | 51.4 |
| $I\bar{4}2m$−C$_3$N$_4$ | 372.0 | 345.8 | 791.9 | 298.8 | 0.1452 | 53.1 |
| $I\bar{4}3d$−C$_3$N$_4$ | 436.6 | 374.4 | 873.4 | 275.2 | 0.1666 | 50.5 |
| $Cmc2_1$-C$_3$N$_4$ | 347.1 | 335.1 | 760.5 | 312.2 | 0.1348 | 54.6 |
| $P4_2/m$-CN | 326.8 | 272.7 | 640.1 | 189.8 | 0.1736 | 40.0 |
| $Pnnm$-CN | 338.6 | 327.0 | 742.2 | 305.0 | 0.1347 | 53.8 |

According to the Oganov model, the five hardest structures in the C-N system are $P\bar{4}n2$-CN$_2$, $I\bar{4}m2$-CN$_2$, $I\bar{4}2d$-CN$_2$, $F\bar{4}3m$-C$_{11}$N$_4$ and $I\bar{4}3d$-C$_3$N$_4$, with hardnesses 87.8, 86.0, 85.6, 83.8 and 83.8 GPa, respectively. Thus, compositions CN$_2$, C$_{11}$N$_4$, and C$_3$N$_4$ are hardness-favored based on the Oganov microscopic model. Strikingly, the hardest structure is $P\bar{4}n2$-CN$_2$, with hardness of 87.8 GPa, just 1.4 GPa less than that of diamond (89.2 GPa based on the Oganov model).

According to the Chen model, the top five hardest structures of the C-N system are $F\bar{4}3m$-C$_{11}$N$_4$, $C2/m$-C$_2$N, $P2_1/c$-C$_2$N, $Cm$-C$_{11}$N$_4$ and $Pmn2_1$-C$_2$N, with hardnesses of 74.1, 73.5, 72.6, 71.5 and 69.6 GPa, respectively. Consequently, compositions C$_{11}$N$_4$ and C$_2$N are hardness-favored based on the Chen macroscopic model. The hardest structure is $F\bar{4}3m$-C$_{11}$N$_4$, with hardness of 74.1 GPa based on the Chen model. It is harder than the well-known superhard cubic-BN, which has a hardness of 62 GPa[39]. This structure is a vacancy-ordered derivative of the diamond structure. The notable feature of the $F\bar{4}3m$-C$_{11}$N$_4$ structure is that its shear modulus is much larger than its bulk modulus, similar to diamond. Besides, this structure has a very small Poisson's ratio (0.0985), which is one of the factors responsible for its superhardness.

Three-dimensional covalent bond network is one of key features of superhard materials. As shown in Figs. 3c, 3d and Supplementary Fig. S6 online, high electron localization is observed near midpoints of C-N, C-C and N-N bonds, indicating infinite three-dimensional covalent bond networks in all these phases. One can also see pronounced maxima of the electron localization function near N atoms, corresponding to the lone electron pair of the nitrogen atom.

The widespread interest in carbon nitrides also arises from their predicted wide band gap, high atomic density and excellent thermal conductivity[18]. The calculated band gaps of the $I\bar{4}2d$-CN$_2$, $P31c$-C$_3$N$_4$, $Cm$-C$_3$N$_4$, $I\bar{4}2m$-C$_3$N$_4$, $I\bar{4}3d$-C$_3$N$_4$, $Cmc2_1$-C$_3$N$_4$, $P4_2/m$-CN and $Pnnm$-CN structures are 3.57, 3.78, 3.74, 2.75, 2.91, 3.43, 3.81 and 3.71 eV, respectively. In all cases, the gap is found to be indirect, except $I\bar{4}2m$-C$_3$N$_4$ (see Supplementary Table S2 online). One should keep in mind that DFT usually underestimates experimental band gaps by ∼ 30%. All of these phases have predicted atomic densities approaching that of diamond (Table 1). On the basis of the high atomic density and bonding topology of these structures, they should be excellent thermal conductors[18].

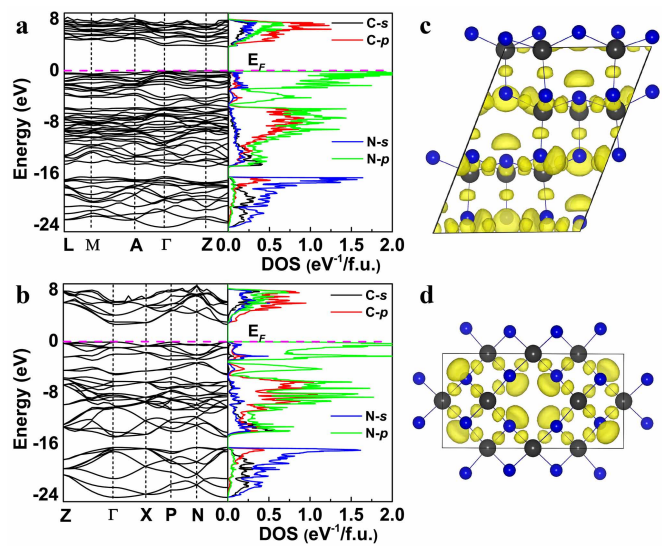

Figure 3 | **Electronic structure of $Cm$-C$_3$N$_4$ and $I\bar{4}2m$-C$_3$N$_4$ at 0 GPa.** Band structure and density of states (DOS) of (A) $Cm$-C$_3$N$_4$, and (B) $I\bar{4}2m$-C$_3$N$_4$. E$_F$, Fermi energy. Electron localization function of (C) $Cm$-C$_3$N$_4$, and (D) $I\bar{4}2m$-C$_3$N$_4$ at ELF = 0.85.

In summary, we have carried out a systematic search for stable phases in the C-N system in the pressure range 0–300 GPa using the *ab initio* evolutionary algorithm USPEX. We have predicted eight stable carbon nitride phases, three of them have never been reported. Carbon nitrides appear as stable phases at pressures above 14 GPa. All carbon nitrides that have stability fields are superhard wide-gap semiconductors or insulators, and remain dynamically and mechanically stable at zero pressure and thus can be quenched to ambient conditions. Among the stable carbon nitrides, the hardest one is $I\bar{4}2d$-CN$_2$ ($Cmc2_1$-C$_3$N$_4$) based on the Oganov (Chen) models. Compositions C$_{11}$N$_4$, C$_2$N, CN$_6$ produce very low-enthalpy metastable states. Considering both stable and low-enthalpy metastable structures, the hardest structure is $P\bar{4}n2$-CN$_2$ ($F\bar{4}3m$-C$_{11}$N$_4$) based on the Oganov (Chen) models, and their hardnesses are close to that of diamond. At ambient pressure the lowest-energy form of C$_3$N$_4$ is $Cc$-C$_3$N$_4$ (s-heptazine), while another structure $Cc$-C$_3$N$_4$ (s-triazine) is a little higher in energy, but both s-heptazine and s-triazine units C$_3$N$_4$ could be synthesized using proper precursors. With respect to the synthesis of superhard materials, $P4_2/m$-CN is a good candidate, because it is thermodynamically stable at the lowest pressure of 14 GPa. In addition, CN$_2$ will be a promising compound to be synthesized in experiment, due to its stability over a wide pressure, and this phase has extremely high hardness.

## Methods

Structure relaxations and energy calculations were performed using density functional theory (DFT)[40] within the generalized gradient approximation (GGA)[41] as implemented in the VASP code[42]. The plane wave kinetic energy cutoff was set to 600 eV. Phonon dispersions were calculated by the supercell approach as implemented in the PHONOPY code[43]. The elastic constants were calculated from the strain-stress relations[44], and hardness was computed using four different models: microscopic Oganov[34,35], Šimůnek[36] and Gao[37] models and the Chen model[38], which is based on macroscopic parameters (elastic moduli). The enthalpy of formation of compounds $C_xN_y$ is calculated by $\Delta H$ (per atom) = $\{H(C_xN_y)-[xH(C) + yH(N)]\}/(x + y)$, where $H(C_xN_y)$, $H(C)$ and $H(N)$ are enthalpies of $C_xN_y$ (per formula), and of stable phases of carbon and nitrogen (per atom), respectively, at given pressure. Energies and phonon dispersion curves $C_3N_4$ structures at zero pressure were checked with the van der Waals corrections[45].

1. Liu, A. Y. & Cohen, M. L. Prediction of new low compressibility solids. *Science* **245**, 841–2 (1989).
2. Wang, X. et al. A metal-free polymeric photocatalyst for hydrogen production from water under visible light. *Nat. Mater.* **8**, 76–80 (2008).
3. Yan, S. C., Li, Z. S. & Zou, Z. G. Photodegradation Performance of g-$C_3N_4$ Fabricated by Directly Heating Melamine. *Langmuir* **25**, 10397–10401 (2009).
4. Bu, Y., Chen, Z., Yu, J. & Li, W. A novel application of g-$C_3N_4$ thin film in photoelectrochemical anticorrosion. *Electrochim. Acta* **88**, 294–300 (2013).
5. Muhl, S. & Méndez, J. M. A review of the preparation of carbon nitride films. *Diam. Relat. Mater.* **8**, 1809–1830 (1999).
6. Solozhenko, V. L., Solozhenko, E. G. & Lathe, C. Synthesis of turbostratic cabon nitride. *J. Superhard Mater.* **24**, 85–86 (2002).
7. Solozhenko, V. L. et al. Equation of state and phase stability of turbostratic carbon nitride. *J. Phys. Chem. Solids* **64**, 1265–1270 (2003).
8. Zhang, M., Wei, Q., Yan, H., Zhao, Y. & Wang, H. A Novel Superhard Tetragonal Carbon Mononitride. *J. Phys. Chem. C* **118**, 3202–3208 (2014).
9. Komatsu, T. Shock synthesis and characterization of new diamond-like carbon nitrides. *Phys. Chem. Chem. Phys.* **6**, 878–880 (2004).
10. Zhou, X.-F. et al. Most likely phase of superhard $BC_2N$ by *ab initio* calculations. *Phys. Rev. B* **76**, 100101 (2007).
11. Li, Q. et al. A novel low compressible and superhard carbon nitride: body-centered tetragonal $CN_2$. *Phys. Chem. Chem. Phys.* **14**, 13081–7 (2012).
12. Mattesini, M. & Matar, S. F. Density-functional theory investigation of hardness, stability, and electron-energy-loss spectra of carbon nitrides with $C_{11}N_4$ stoichiometry. *Phys. Rev. B* **65**, 075110 (2002).
13. Tian, F. et al. Superhard semiconducting $C_3N_2$ compounds predicted via first-principles calculations. *Phys. Rev. B* **78**, 235431 (2008).
14. Oganov, A. R. & Glass, C. W. Crystal structure prediction using *ab initio* evolutionary techniques: Principles and applications. *J. Chem. Phys.* **124**, 244704 (2006).
15. Oganov, A. R., Lyakhov, A. O. & Valle, M. How Evolutionary Crystal Structure Prediction Works and Why. *Acc. Chem. Res.* **44**, 227–237 (2011).
16. Lyakhov, A. O., Oganov, A. R., Stokes, H. T. & Zhu, Q. New developments in evolutionary structure prediction algorithm USPEX. *Comput. Phys. Commun.* **184**, 1172–1182 (2013).
17. Oganov, A. R., Ma, Y., Lyakhov, A. O., Valle, M. & Gatti, C. Evolutionary Crystal Structure Prediction as a Method for the Discovery of Minerals and Materials. *Rev. Mineral. Geochem.* **71**, 271–298 (2010).
18. Teter, D. M. & Hemley, R. J. Low-Compressibility Carbon Nitrides. *Science* **271**, 53–55 (1996).
19. Fang, L., Ohfuji, H., Shinmei, T. & Irifune, T. Experimental study on the stability of graphitic $C_3N_4$ under high pressure and high temperature. *Diam. Relat. Mater.* **20**, 819–825 (2011).
20. Kojima, Y. & Ohfuji, H. Structure and stability of carbon nitride under high pressure and high temperature up to 125 GPa and 3000 K. *Diam. Relat. Mater.* **39**, 1–7 (2013).
21. He, J. et al. Predicting hardness of dense $C_3N_4$ polymorphs. *Appl. Phys. Lett.* **88**, 101906 (2006).
22. Zhang, Y., Sun, H. & Chen, C. Strain dependent bonding in solid $C_3N_4$: High elastic moduli but low strength. *Phys. Rev. B* **73**, 064109 (2006).
23. Wang, X., Blechert, S. & Antonietti, M. Polymeric Graphitic Carbon Nitride for Heterogeneous Photocatalysis. *ACS Catal.* **2**, 1596–1606 (2012).
24. Kroke, E. gt-$C_3N_4$ The First Stable Binary Carbon(IV) Nitride. *Angew. Chem. Int. Ed.* **53**, 11134–11136 (2014).
25. Kroke, E. et al. Tri-s-triazine derivatives. Part I. From trichloro-tri-s-triazine to graphitic $C_3N_4$ structures. *New J. Chem.* **26**, 508–512 (2002).
26. Wei, W. & Jacob, T. Strong excitonic effects in the optical properties of graphitic carbon nitride g-$C_3N_4$ from first principles. *Phys. Rev. B* **87**, 085202 (2013).
27. Deifallah, M., McMillan, P. F. & Corà, F. Electronic and Structural Properties of Two-Dimensional Carbon Nitride Graphenes. *J. Phys. Chem. C* **112**, 5447–5453 (2008).
28. Bojdys, M. J., Mller, J.-O., Antonietti, M. & Thomas, A. Ionothermal Synthesis of Crystalline, Condensed, Graphitic Carbon Nitride. *Chem. Eur. J.* **14**, 8177–8182 (2008).
29. Gracia, J. & Kroll, P. Corrugated layered heptazine-based carbon nitride: the lowest energy modifications of $C_3N_4$ ground state. *J. Mater. Chem.* **19**, 3013–3019 (2009).
30. Zelisko, M. et al. Anomalous piezoelectricity in two-dimensional graphene nitride nanosheets. *Nat. Commun.* **5**, 4284 (2014).
31. Wang, X. Polymorphic phases of $sp^3$-hybridized superhard CN. *J. Chem. Phys.* **137**, 184506 (2012).
32. Zhu, Q., Feyac, O. D., Boulfelfela, S. E. & Oganova, A. R. Metastable Host-Guest Structure of Carbon. *J. Superhard Mater.* **36**, 246–256 (2014).
33. Horvath-Bordon, E. et al. High-Pressure Synthesis of Crystalline Carbon Nitride Imide, $C_2N_2$ (NH). *Angew. Chem. Int. Ed.* **46**, 1476–1480 (2007).
34. Lyakhov, A. O. & Oganov, A. R. Evolutionary search for superhard materials: Methodology and applications to forms of carbon and $TiO_2$. *Phys. Rev. B* **84**, 092103 (2011).
35. Li, K., Wang, X., Zhang, F. & Xue, D. Electronegativity Identification of Novel Superhard Materials. *Phys. Rev. Lett.* **100**, 235504 (2008).
36. Šimůnek, A. & Vackář, J. Hardness of Covalent and Ionic Crystals: First-Principle Calculations. *Phys. Rev. Lett.* **96**, 085501 (2006).
37. Gao, F. et al. Hardness of Covalent Crystals. *Phys. Rev. Lett.* **91**, 015502 (2003).
38. Chen, X.-Q., Niu, H., Li, D. & Li, Y. Modeling hardness of polycrystalline materials and bulk metallic glasses. *Intermetallics* **19**, 1275–1281 (2011).
39. Solozhenko, V. L., Andrault, D., Fiquet, G., Mezouar, M. & Rubie, D. C. Synthesis of superhard cubic $BC_2N$. *Appl. Phys. Lett.* **78**, 1385–1387 (2001).
40. Blöchl, P. E. Projector augmented-wave method. *Phys. Rev. B* **50**, 17953–17979 (1994).
41. Perdew, J. P., Burke, K. & Ernzerhof, M. Generalized Gradient Approximation Made Simple. *Phys. Rev. Lett.* **77**, 3865–3868 (1996).
42. Kresse, G. & Furthmüller, J. Efficient iterative schemes for *ab initio* total-energy calculations using a plane-wave basis set. *Phys. Rev. B* **54**, 11169–11186 (1996).
43. Togo, A., Oba, F. & Tanaka, I. First-principles calculations of the ferroelastic transition between rutile-type and $CaCl_2$-type $SiO_2$ at high pressures. *Phys. Rev. B* **78**, 134106 (2008).
44. Le Page, Y. & Saxe, P. Symmetry-general least-squares extraction of elastic data for strained materials from *ab initio* calculations of stress. *Phys. Rev. B* **65**, 104104 (2002).
45. Grimme, S. Semiempirical GGA-type density functional constructed with a long-range dispersion correction. *J. Comp. Chem.* **27**, 1787–1799 (2006).

## Acknowledgments

We thank the National Science Foundation (EAR-1114313, DMR-1231586), DARPA (Grants No. W31P4Q1210008 and No. W31P4Q1310005), the Government (No.14. A12.31.0003) of Russian Federation for financial support, and Foreign Talents Introduction and Academic Exchange Program (No. B08040). Calculations were carried out in part at the Center for Functional Nanomaterials, Brookhaven National Laboratory, which is supported by the U.S. Department of Energy, Office of Basic Energy Sciences, under Contract No. DE-SC0012704. We also thank Purdue University Teragrid and TACC Stampede system for providing computational resources and technical support for this work (Charge No. TG-DMR110058).

## Author contributions

H.F.D performed all the calculations presented in this article with help from Q.Z. and G.R.Q. Research was designed by A.R.O. H.F.D. and A.R.O. wrote the first draft of the paper and all authors contributed to revisions.

## Additional information

**Supplementary information** accompanies this paper at http://www.nature.com/scientificreports

**Competing financial interests:** The authors declare no competing financial interests.

**How to cite this article:** Dong, H., Oganov, A.R., Zhu, Q. & Qian, G.-R. The phase diagram and hardness of carbon nitrides. *Sci. Rep.* **5**, 9870; DOI:10.1038/srep09870 (2015).